# The Stock Price Relationship between Holding Companies and Subsidiaries: A Case study of Indonesia Multiholding Companies

**Muhammad Aufaristama[1,2]**

[1] Department of Applied Earth Sciences, Faculty Geo-information Sciences and Earth Observation, University of Twente, the Netherlands
[2] Independent Individual Investor of Indonesian Stock Exchange

*This study aimed to examine the correlation between the stock prices of two major Indonesian holding companies, MNC Group and Elang Mahkota Teknologi (Emtek) Group, and their respective subsidiaries as case studies. The data for the analysis were collected from 2013 to 2022, and Spearman correlation was used to determine the strength and direction of the relationship between the stock prices of the holding companies and their subsidiaries. The results of the analysis revealed that there were varying degrees of correlation between the stock prices of the holding companies and their subsidiaries. The strongest positive correlation was observed between BHIT and BMTR, while the weakest correlations were found between BHIT and IPTV, and BHIT and MSIN. The correlations were also found to have changed over time, possibly due to market conditions, company-specific events, or changes in industry sectors.In the case of Emtek Group, the analysis suggested that EMTK's stock price movements had a significant impact on the stock prices of its subsidiaries, with varying strengths of relationships. The negative correlation between EMTK and SCMA over the entire period suggested an inverse relationship, while positive correlations with BUKA, AMOR, BBHI, and RSGK indicated a tendency to move in the same direction as EMTK's stock price. The correlations were found to have increased over time, possibly due to market conditions and EMTK's ownership stake in these companies. Overall, the findings of this study suggest that there is a complex interplay between the stock prices of parent companies and their subsidiaries, and that there are a variety of factors that can influence these relationships over time. These findings may be useful for investors in making informed decisions about their investment portfolios, as changes in the correlations could impact their portfolio's performance.*

**Keywords:** stock prices; holding companies; subsidiaries; Indonesian multiholding companies; spearman correlation

**JEL Classification:** G10, G110

Corresponding author's email: m.aufaristama@utwente.nl

# Introduction

Multinational corporations have become increasingly significant players in the global economy in recent years, with a growing presence in many countries, including Indonesia (Dunning, 1993; Rugman, 2012). Their stocks are traded on various national and international stock exchanges (Ogawa & Zhang, 2017). Holding companies, in particular, are a type of multinational corporation that owns and controls one or more companies, known as subsidiaries (Sharma & Lawrence, 2016). A holding company is a type of business entity that is commonly organized as a corporation or limited liability company (LLC) (Stapledon, 2019). Holding companies are primarily characterized by their role as a passive investor and do not engage in manufacturing, product or service sales, or other operational activities (Bulow & Geanakoplos, 2014). Instead, they primarily hold controlling ownership in other companies which operate in various industries (Hwang & Chung, 2020). One important aspect of the relationship between holding companies and their subsidiaries is the impact on stock prices. The stock prices of holding companies and their subsidiaries are interconnected, and understanding this relationship is essential for investors to make informed investment decisions.

The stock prices of holding companies and their subsidiaries can be influenced by various factors, including the overall economic climate and the performance of individual companies. In Indonesia, the steady growth of the economy has been shown to have a positive impact on the stock prices of many companies, including holding companies and their subsidiaries (Fitriany & Prasetyo, 2018; Mustikasari, Suhadak, & Budiarti, 2021). However, economic downturns can have a negative impact on stock prices, affecting both holding companies and their subsidiaries (Mahendra & Hermanto, 2019). The performance of individual companies can also play a significant role in the stock prices of holding companies and their subsidiaries. A strong performance by a subsidiary can have a positive impact on the stock price of the holding company, while poor performance can have a negative impact (Puspitasari & Wardhani, 2020). Regulatory changes can also have a significant impact on the stock prices of holding companies and their subsidiaries. In Indonesia, the government has implemented various regulatory changes in recent years, such as changes to tax laws and labor laws. These changes can affect the profitability of companies, and in turn, impact their stock prices. A study by Nurlaela and Wirjodirdjo (2020) found that regulatory changes had a significant impact on the stock prices of firms listed on the Indonesia Stock Exchange (IDX). The study also suggests that regulatory changes can have a differential impact on the stock prices of different firms, depending on their industry and size. Therefore, it is important for investors to monitor regulatory changes and their potential impact on the companies in which they invest.

A positive relationship between the stock prices of holding companies and their subsidiaries has been suggested in previous studies by Li and Zhang (2019) and Soares and Matos (2015), as changes in one may affect the other. However, the full extent of this relationship and the contributing factors remain unclear. Specifically, in the context of Indonesian multiholding companies, little research has been conducted on the stock price relationship between holding companies and their subsidiaries. The stock price relationship between holding companies and their subsidiaries in Indonesia, with a focus on multiholding companies, is the main focus of this study. By examining this relationship, insights into the dynamics of these entities and their impact on the wider stock market can be gained. Relevant literature for this study has been provided by prior research conducted by Huang (2018) and Rahman and Kim (2020).

## Literature Review

The relationship between the stock prices of holding companies and their subsidiaries has been the subject of extensive research. While some studies suggest a positive relationship between the two, others suggest a negative relationship. However, most studies agree that there is a significant relationship between the stock prices of holding companies and their subsidiaries. For instance, Lins and Servaes (1999) found a positive relationship between the stock prices of holding companies and their subsidiaries in their study of European companies. Similarly, Hong and Lobo (2004) found a positive relationship between the stock prices of holding companies and their subsidiaries in their study of US companies. De Franco, Kothari, and Verdi (2011) examine the stock price reactions of parent companies and their subsidiaries in response to the release of financial statements. They find a positive correlation between the stock prices of parent companies and their subsidiaries, indicating that investors perceive the financial performance of both entities to be closely related. Chen and Wu (2013) analyzed the stock price reactions of listed holding companies in China when they announced equity carve-outs of their subsidiaries. They found that the stock prices of both the holding company and the subsidiary experience positive abnormal returns following the announcement, suggesting that equity carve-outs increase the value of both entities. Similarly, Serra, de la Fuente, and Garcia-Lacalle (2018) investigated the stock price relationship between holding companies and their subsidiaries in Spain. Their study finds a positive relationship between the stock prices of holding companies and their subsidiaries, indicating that investors view both entities as closely related and interdependent. Finally, Fang et al (2017) examined the stock price reactions of listed companies in China that engaged in cross-border mergers and acquisitions. They found that the stock prices of both the acquiring company and the target company experience positive abnormal returns following the announcement, indicating that investors perceive the two entities as closely related and expect the merger or acquisition to create value for both.

The correlation between stock prices has been a widely researched topic in finance, as investors seek to understand the relationships between different stocks in their portfolios. One common method used to measure the correlation between stock prices is the Spearman correlation coefficient. For example, studies by Akhtar et al (2018) and Rehman et al (2019) use Spearman correlation to examine the relationships between stock prices in the Pakistani stock market. Both studies find significant positive correlations between the stock prices of companies in the same sector, suggesting that investors tend to group companies in similar industries together in their investment decisions. Similarly, a study by Farooq and Nasir (2019) uses Spearman correlation to examine the relationships between stock prices in the Indian stock market. The study finds significant positive correlations between the stock prices of companies in the same industry, indicating that investors tend to view companies in similar businesses as closely related. Another study by Kim et al (2021) uses Spearman correlation to investigate the relationships between stock prices in the Korean stock market. The study finds significant positive correlations between the stock prices of companies in the same sector and significant negative correlations between the stock prices of companies in different sectors, suggesting that investors consider sector-specific factors when making investment decisions. The studies indicate that Spearman correlation is an effective statistical tool for investigating the relationships between stock prices across diverse markets and industries, and it will be utilized in the present research.

## Data and Methods

*Datasets*

The datasets were obtained comprising daily closing stock prices for two holding companies, MNC Asia Holding (BHIT) and Elang Mahkota Teknologi (EMTK) Group, as well as their respective subsidiaries from the IDX. The data covers a period of ten years, starting from January 1, 2013, and ending on December 31, 2022. The selection of these companies was based on several factors. Firstly, the companies were chosen because they are holding companies listed in the IDX-IC C311 sector, which is the Multi-Sector Holding Companies sector IDX (IDX,2021) This sector includes companies that operate in various industries, and the selection of holding companies from this sector could provide a diversified perspective on the performance of the Indonesian stock market. Secondly, the companies were selected because both have more than five subsidiaries listed in IDX. The inclusion of subsidiaries in the datasets can provide a more comprehensive view of the performance of the holding companies. Lastly, the stocks selected also have similar subsidiaries in the sectors of media entertainment, financial, and hospitality/or healthcare. This criterion was used to ensure that the selected companies had some level of similarity in their operations and could be compared to each other in terms of their correlation with respective subsidiaries. Table 1 includes a list of the holdings and subsidiaries' names, their corresponding stock symbols and their respective business sectors.

Table 1: List of the holdings and subsidiaries' names, their corresponding stock symbols and their respective business sectors

| Holding | Subsidiaries | Sectors |
|---|---|---|
| MNC Asia Holding (BHIT) | Global Mediacom(BMTR) | Sub-holding, multisectors |
| | Media Nusantara Citra (MNCN) | Media & Entertainment |
| | MNC Digital Entertainment (MSIN) | Media & Entertainment |
| | MNC Vision Networks (IPTV) | Media & Entertainment |
| | MNC Sky Vision (MSKY) | Media & Entertainment |
| | MNC Kapital Indonesia (BCAP) | Finance |
| | Bank MNC Internasional (BABP) | Finance |
| | MNC Land (KPIG) | Hospitality/Healthcare |
| | MNC Energy Investments (IATA) | Hospitality/Healthcare |
| Elang Mahkota Teknologi (EMTK | Surya Citra Media (SCMA) | Media & Entertainment |
| | Bukalapak (BUKA) | Consumer |
| | Ashmore Asset Management Indonesia (AMOR) | Finance |
| | Allo Bank Indonesia (BHBI) | Finance |
| | Sarana Meditama Metropolitan (SAME) | Hospitality/Healthcare |
| | Kedoya Adyaraya (RSGK) | Hospitality/Healthcare |

*Methods*

The Spearman correlation coefficient was utilized to assess the daily closing stock prices of each holding company and its subsidiaries. Mohanty and Rabi (2012) conducted an empirical analysis of stock prices using the correlation coefficient and found that it is a useful tool for predicting the behavior of stock prices. The coefficient is particularly effective when the relationship between the two variables is non-linear, and it can provide insights into how one stock's price might affect the other. The Spearman correlation coefficient works by ranking the daily closing prices of each holding company and its subsidiaries, then calculating the correlation between the rankings. This method is resilient to outliers, which can skew the results of other types of correlation coefficients (Kim et al., 2018). The calculation of the Spearman correlation coefficient can be expressed using the formula (1) provided in the original text, as well as other statistical methods described in textbooks such as Wackerly et al. (2014) and Montgomery et al. (2012)

$$r_s = 1 - \frac{6 \sum d_i^2}{N(N^2+1)} \tag{1}$$

Where $d_i$ is the difference between ranks for each data pair, and $N$ is the number of data pairs.

The resulting coefficient ranges from -1 to 1, with a positive value indicating that the two stocks are positively correlated, a negative value indicating that they are negatively correlated, and a value of zero indicating that there is no correlation. This information can be used to make predictions about the behavior of stock prices, as explained in the original text. Overall, the Spearman correlation coefficient is a reliable and widely-used tool for analyzing stock market price movements, and its effectiveness has been demonstrated in both empirical studies and theoretical analyses. In this study, the Spearman correlation was used to explore the relationship between stock prices over the past ten, five, and three years. To provide clarity, the normalization of stock prices for the MNC and Emtek group is displayed in Figure 1.

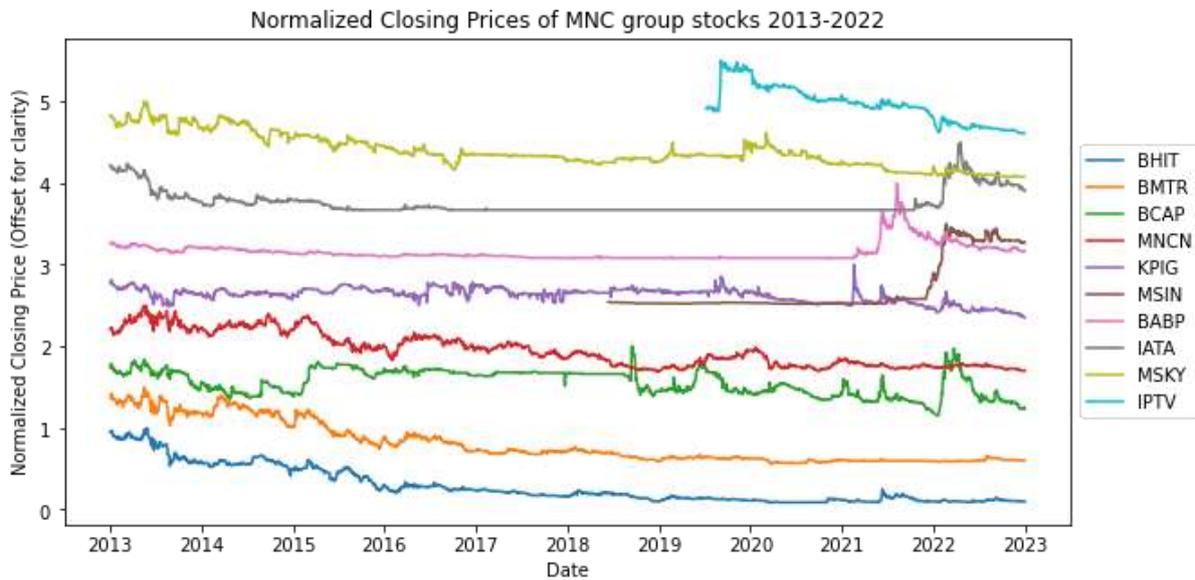

(a)

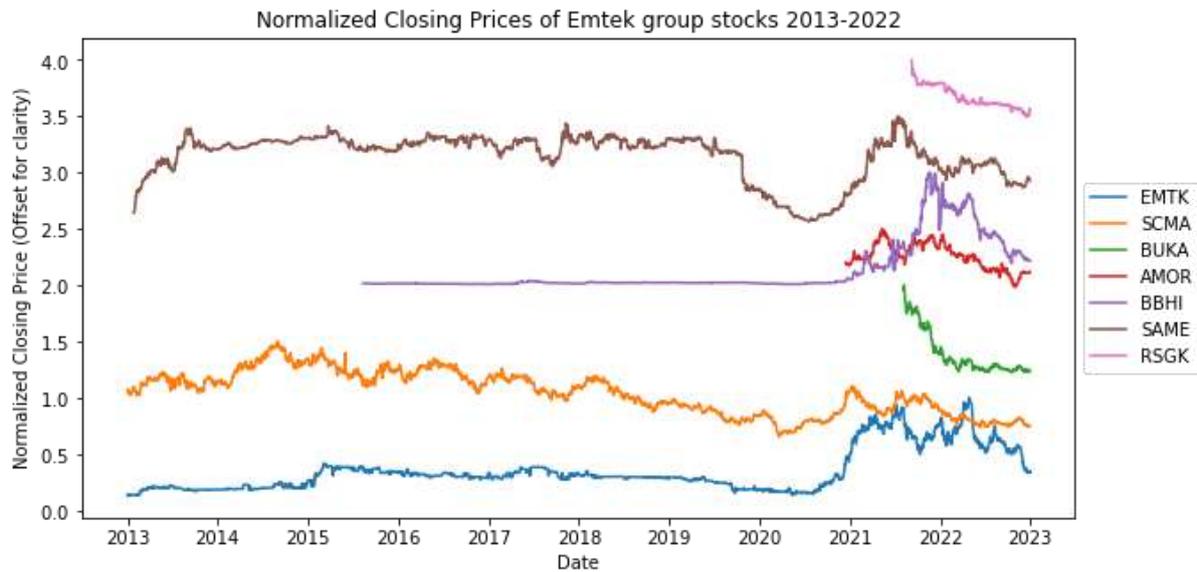

(b)

Figure 1 (a) Normalization of stock prices of MNC group (offset for clarity); (b) Normalization of stock prices of Emtek group (offset for clarity)

## Results and Discussions

*MNC group stock price correlation*

The Spearman correlation coefficients matrix in Figure 2a demonstrate the relationships between the stock prices of MNC group and its subsidiaries from 2013 to 2022. BHIT and BMTR exhibit a strong positive correlation (r = 0.96), indicating that their stock prices tend to move in the same direction. Similarly, BHIT and MNCN have a moderately strong positive correlation (r = 0.88), suggesting that their stock prices are somewhat related. However, BHIT's correlations with BCAP (r = 0.5), KPIG (r = 0.51), and IATA (r = 0.41) are weaker, implying that there may be less of a relationship between the stock prices of these companies and BHIT. BHIT and MSIN have a very weak correlation (r = 0.13), suggesting that there may be little to no relationship between their stock prices. The correlation between BHIT and BABP is also weak (r = 0.36), albeit slightly stronger than the correlation with MSIN. On the other hand, BHIT and MSKY exhibit a moderately strong correlation (r = 0.76), indicating that their stock prices tend to move together. Lastly, the correlation between BHIT and IPTV is very weak (r = 0.012), suggesting that there is little to no relationship between their stock prices.

From 2018 to 2022, the Spearman correlation coefficients demonstrate the relationship between BHIT and its subsidiaries' stock prices (Figure 2b). The correlation between BHIT and BMTR is 0.79, which indicates that their stock prices are positively correlated and tend to move in the same direction. The correlation between BHIT and MNCN is 0.26, suggesting that their stock prices are somewhat related but have a weaker correlation compared to BHIT and BMTR. BHIT and BCAP have a correlation of 0.42, which implies that there is a moderate positive relationship between their stock prices. BHIT and KPIG have a correlation of 0.55, indicating a moderately strong positive relationship between their stock prices. The correlation between BHIT and MSIN is 0.13, implying that there may be little to no relationship between their stock prices. BHIT and MSKY have a correlation of 0.038, which is weak and suggests that their stock prices are not strongly related. The correlation between BHIT and BABP is 0.016, indicating that their stock prices are weakly correlated. The correlation between BHIT and IPTV is 0.012, suggesting little to no relationship between their stock prices. Finally, BHIT and IATA have a negative correlation of -0.32, indicating an inverse relationship between their stock prices.

In Figure 2c, from 2020 to 2022, the Spearman correlation coefficients demonstrate the relationship between BHIT and its subsidiaries' stock prices. The correlation between BHIT and BMTR is 0.61, indicating that their stock prices are positively correlated, but the correlation is weaker than in the previous period. The correlation between BHIT and MNCN is 0.07, suggesting that there is little to no relationship between their stock prices. BHIT and BCAP have a negative correlation of -0.17, indicating an inverse relationship between their stock prices. BHIT and KPIG have a correlation of 0.02, suggesting that there is little to no relationship between their stock prices. The correlation between BHIT and MSIN is 0.4, implying a moderate positive relationship between their stock prices. BHIT and MSKY have a correlation of -0.39, indicating an inverse relationship between their stock prices. The correlation between BHIT and BABP is 0.53, indicating a moderate positive relationship between their stock prices. The correlation between BHIT and IPTV is -0.22, suggesting an inverse relationship between their stock prices. Finally, BHIT and IATA have a correlation of 0.078, indicating that there is little to no relationship between their stock prices.

In overall, the strongest positive correlation was observed between BHIT and BMTR, which is not surprising considering BHIT owns a significant portion of BMTR's shares (45.75%) according to Q3 2022 financial statement. Similarly, there was a strong positive correlation between BHIT and MNCN, which is likely due to BMTR's majority ownership of MNCN.

On the other hand, the weakest correlations were found between BHIT and IPTV, and BHIT and MSIN, indicating that these companies' stock prices may be less influenced by BHIT's performance. Furthermore, the analysis suggests that the correlation coefficients between BHIT and its subsidiaries' stock prices have changed over time, which could be influenced by various factors such as market conditions, company-specific events, or changes in the industry sectors. For example, the negative correlation between BHIT and IATA during the 2020-2022 period may indicate that BHIT's performance is inversely related to IATA's change business sector from transportation to energy.

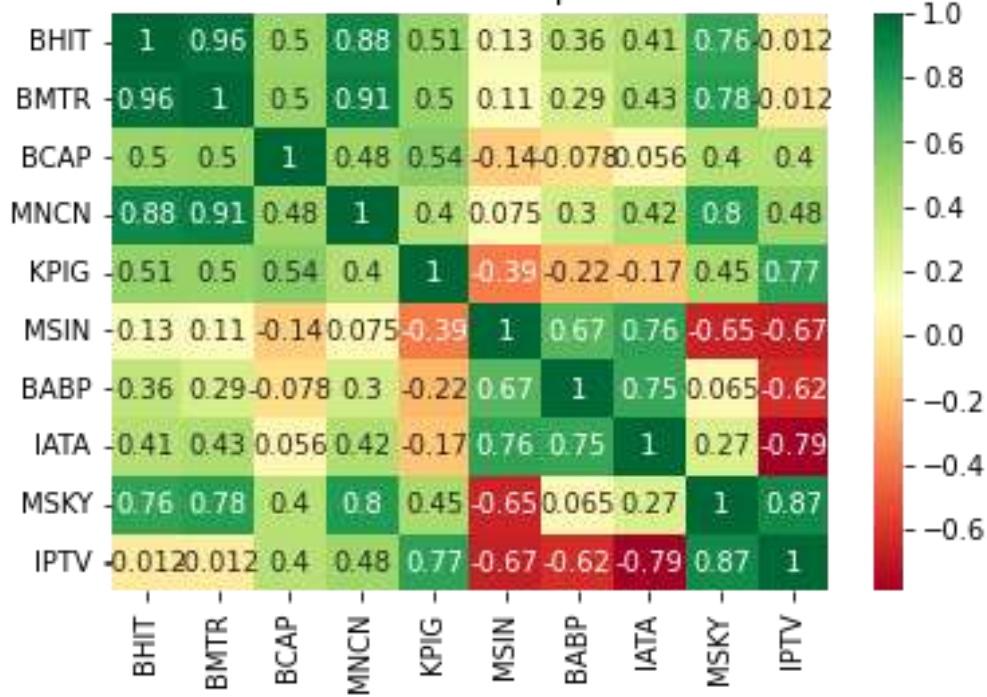

(a)

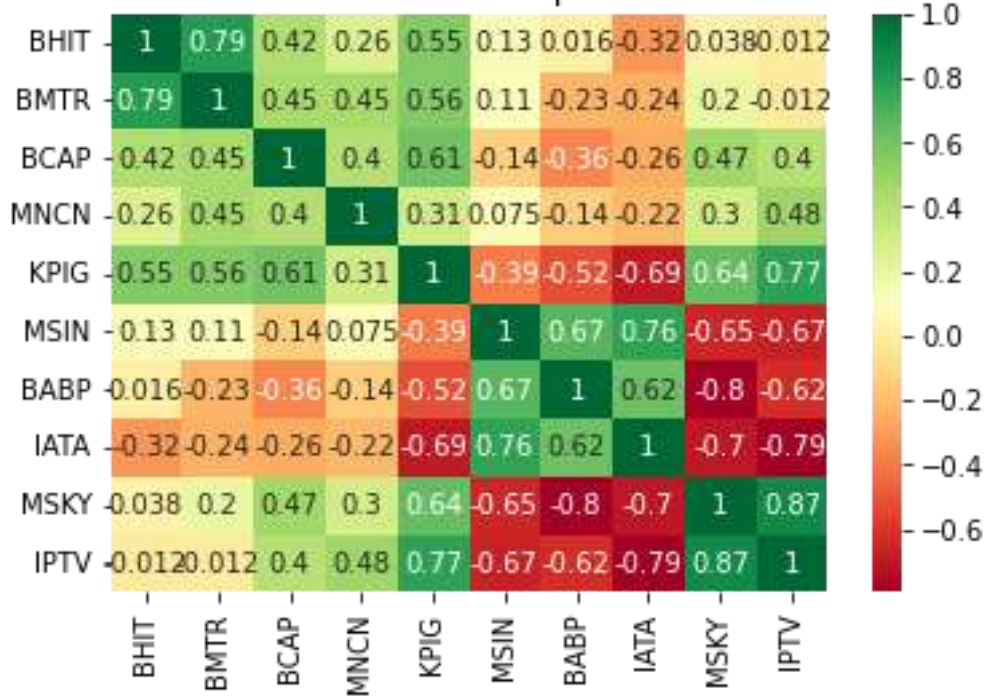

(b)

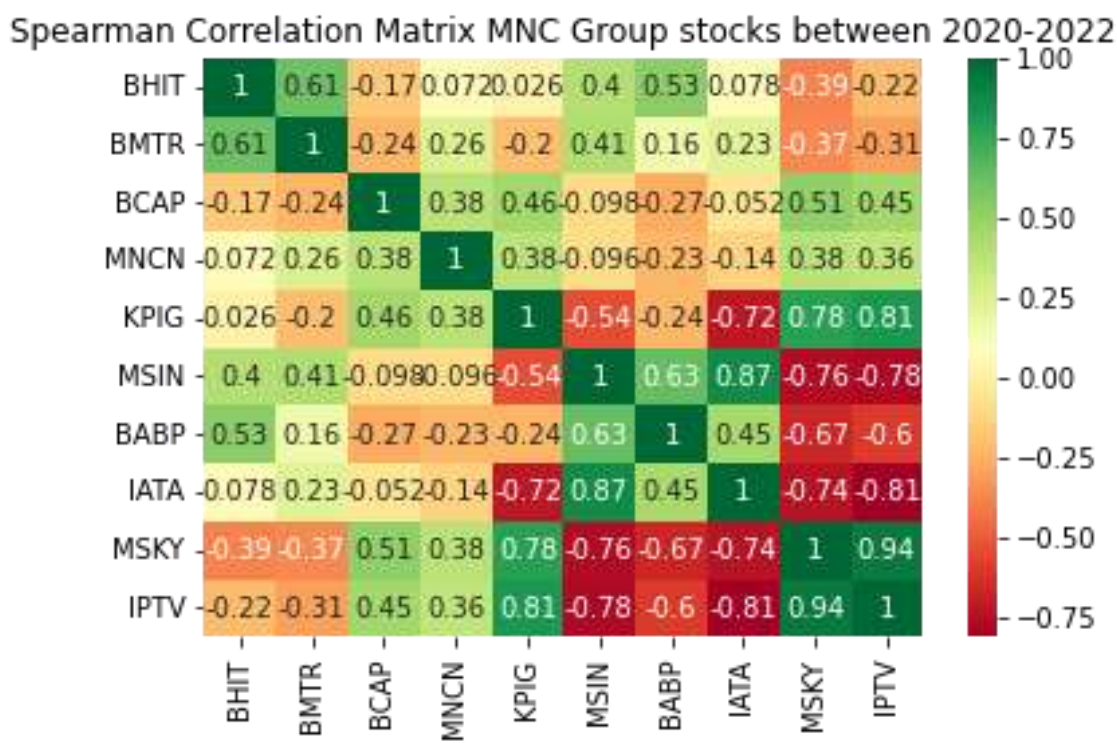

(c)

Figure 2. Spearman correlation matrix for MNC group stocks price between (a) 2013-2022; (b) 2018-2022; and (c) 2020-2022

*Emtek group stock price correlation*

Figures 3a and 3b display the correlation coefficients of Emtek group's stock prices and those of its subsidiaries from 2013 to 2022 and specifically from 2018 to 2022. Over the entire period, EMTK's negative correlation with SCMA (-0.55) suggests that when EMTK's stock price falls, SCMA's stock price tends to rise and vice versa. On the other hand, positive correlations between EMTK and BUKA (0.56), AMOR (0.51), BBHI (0.61), and RSGK (0.61) indicate that the stock prices of these companies tend to move in the same direction as EMTK's stock price. Notably, BUKA had an initial public offering in 2021, as did RSGK, and AMOR had an IPO in 2020. These events may have contributed to the observed high correlations.

From 2018 to 2022, the positive correlation between EMTK and SCMA (0.56) increased, suggesting a moderate to strong relationship between the two companies, likely influenced by similar market conditions during the China-U.S. trade war in early 2019 that impacted IDX and the COVID-19 pandemic. Additionally, the high positive correlations between EMTK and BUKA (0.56), AMOR (0.51), SAME (0.32), and RSGK (0.61) suggest that the stock prices of these companies move in the same direction as EMTK's stock price, possibly due to EMTK's significant ownership stake in these companies.

The very high positive correlation between EMTK and SAME (0.86) from 2020 to 2022 (Figure 3c) suggests a very strong relationship between the two companies, likely due to EMTK's ownership stake in SAME or SAME's significant contribution to the acquisition of RSGK. The high positive correlation between EMTK and BBHI (0.85) from 2013 to 2022 and 0.7 from 2020 to 2022 indicates a strong relationship between the two companies, likely influenced by EMTK's significant ownership stake in BBHI, which operates in the digital banking industry.

In summary, the analysis of the correlations between EMTK's stock prices and those of its subsidiaries provides insights into the relationships among these companies, indicating that EMTK's stock price movements have a positive or negative impact on the stock prices of its subsidiaries, depending on the strength of the relationship. IPOs, market conditions, and significant ownership stakes may also contribute to the observed correlations.

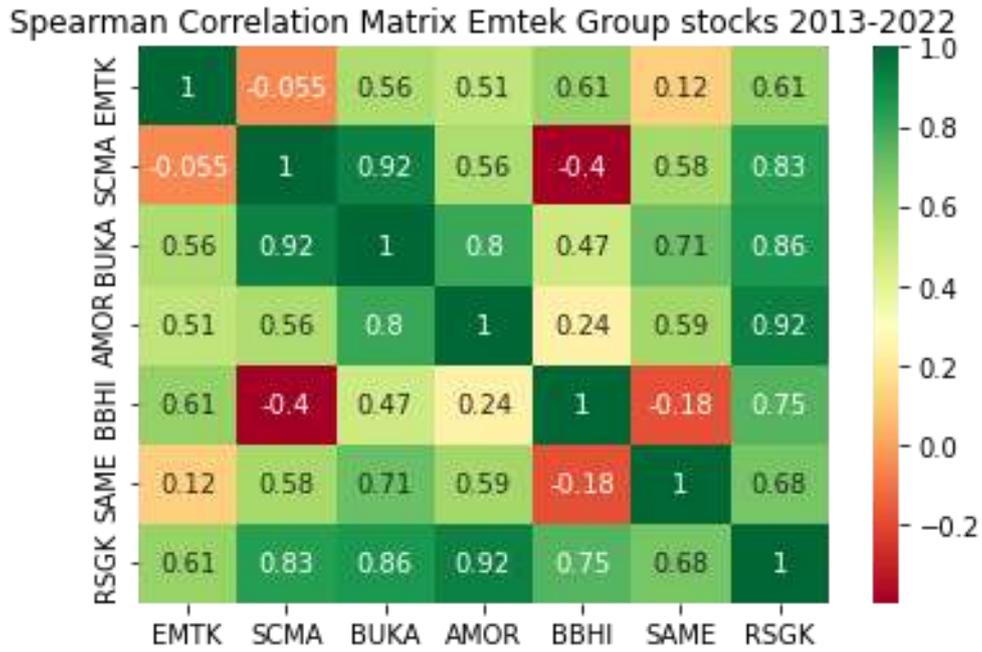

(a)

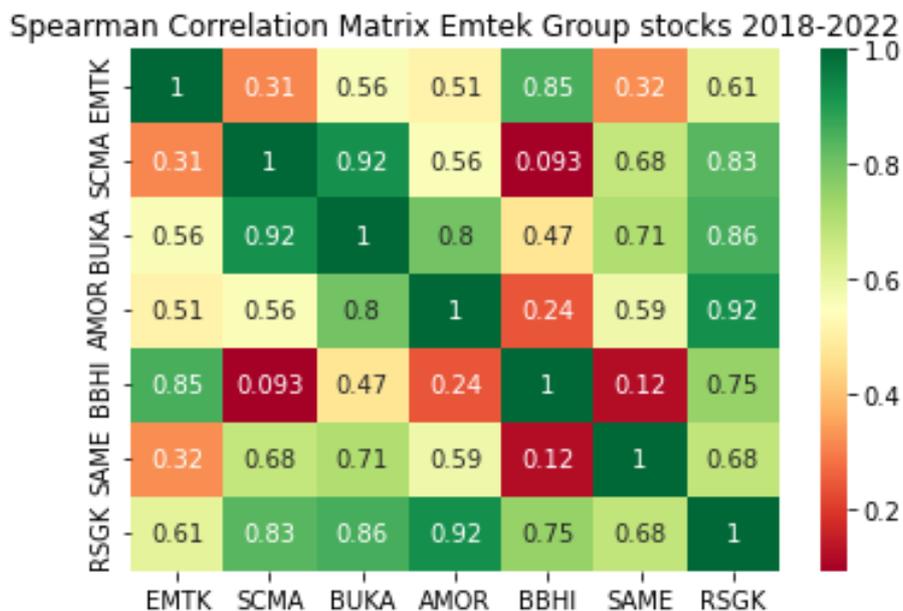

(b)

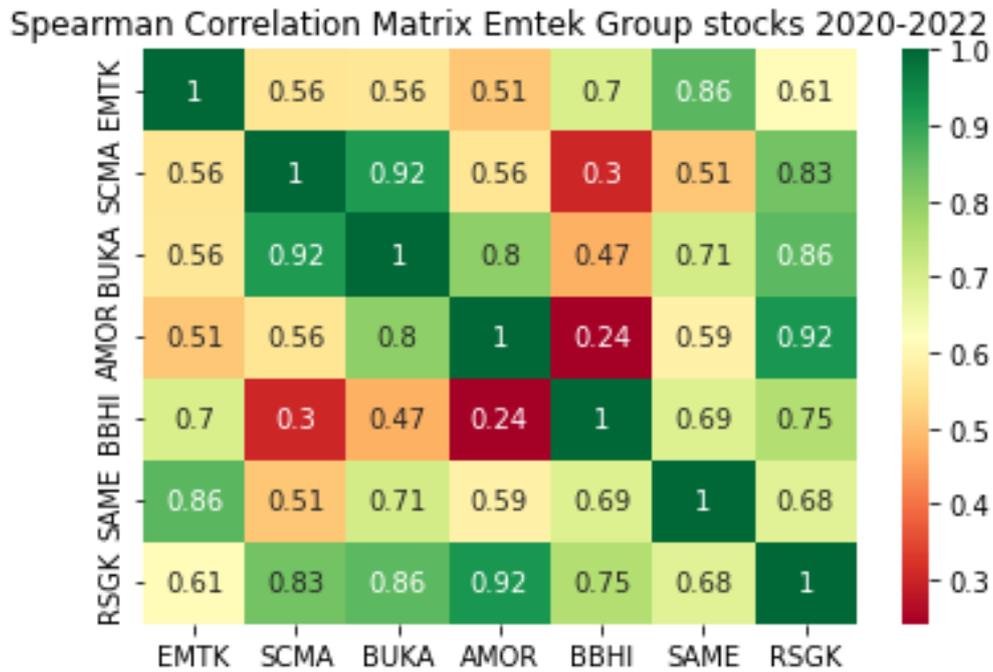

(c)

Figure 3. Spearman correlation matrix for Emtek group stocks price between (a) 2013-2022; (b) 2018-2022; and (c) 2020-2022

## Conclusions

Based on the results, it can be concluded that there are varying degrees of correlation between the stock prices of MNC Group and its subsidiaries. The strongest positive correlation was observed between BHIT and BMTR, and the weakest correlations were found between BHIT and IPTV, and BHIT and MSIN. It is worth noting that BHIT owns a significant portion of BMTR's shares (45.75%), which may explain the strong correlation between their stock prices. Similarly, the majority ownership of MNCN by BMTR may explain the moderately strong correlation between BHIT and MNCN. In contrast, the weakest correlations were found between BHIT and IPTV, and BHIT and MSIN, indicating that these companies' stock prices may be less influenced by BHIT's performance. The analysis also suggests that the correlation coefficients between BHIT and its subsidiaries' stock prices have changed over time, which could be influenced by various factors such as market conditions, company-specific events, or changes in industry sectors. Furthermore, the negative correlation between BHIT and IATA during the 2020-2022 period may indicate that BHIT's performance is inversely related to IATA's change in business sector from transportation to energy.

Regarding Emtek Group's stock price correlations with its subsidiaries indicates that EMTK's stock price movements have a significant impact on the stock prices of its subsidiaries, with varying strengths of relationships. The negative correlation between EMTK and SCMA over the entire period suggests an inverse relationship, while positive correlations with BUKA, AMOR, BBHI, and RSGK indicate a tendency to move in the same direction as EMTK's stock price. The correlations have increased over time, likely due to market conditions and EMTK's ownership stake in these companies. IPOs and market conditions may also play a role in the observed correlations.

The correlation analysis provides valuable insights into the relationships between holding companies and its subsidiaries' stock prices. The results suggest that some subsidiaries may have a stronger relationship with holding companies than others and that these relationships can change over time. This information may be useful for investors in making informed decisions about their investment portfolios. Investors should consider these varying degrees of correlation when making decisions regarding their investment portfolios, as changes in the correlations could impact their portfolio's performance

Overall, these findings suggest that there is a complex interplay between the stock prices of parent companies and their subsidiaries, and that there are a variety of factors that can influence these relationships over time.

## Acknowledgements

I would like to express our gratitude to the Indonesian Stock Market (IDX) for providing us with access to the financial statements of companies used in this study. I would also like to thank Yahoo Finance for providing the historical stock price movements of the companies.